\documentclass[preprint,12pt]{elsarticle}

\usepackage{amssymb,amsmath}
\usepackage{amsthm}
\newtheorem{theorem}{Theorem}
\newtheorem{lemma}{Lemma}
\newtheorem{proposition}{Proposition}
\newtheorem{example}{Example}
\newtheorem{definition}{Definition}
\newtheorem{corollary}{Corollary}
\newtheorem{conjecture}{Conjecture}
\newtheorem{remark}{Remark}

\journal{European Journal of Combinatorics}

\begin{document}
\begin{frontmatter}


\author{Anna E. Frid}
\ead{anna.e.frid@gmail.com}
\address{Aix Marseille Univ, CNRS, Centrale Marseille, I2M, Marseille, France}

\title{Sturmian numeration systems and decompositions to palindromes}




\begin{abstract}
We extend classical Ostrowski numeration systems, closely related to Sturmian words, by allowing a wider range of coefficients, so that possible representations of a number $n$ better reflect the structure of the associated characteristic Sturmian word. In particular, this extended numeration system helps to catch occurrences of palindromes in a characteristic Sturmian word and thus to prove for Sturmian words the following conjecture stated in 2013 by Puzynina, Zamboni and the author: If a word is not periodic, then for every $Q>0$ it has a prefix which cannot be decomposed to a concatenation of at most $Q$ palindromes.
\end{abstract}


\end{frontmatter}
\section{Introduction}
Ostrowski numeration systems, first introduced in 1921 \cite{ostr}, are closely related to continued fractions. A classical example of an Ostrowski numeration system is the Fibonacci (or Zeckendorf) numeration system, first described by Lekkerkerker in 1952 \cite{lekkerkerker,zeckendorf}, where a number is represented as a sum of Fibonacci numbers, but not consecutive ones (since the sum of two consecutive Fibonacci numbers is the next Fibonacci number). In some studies, the idea is extended to allow consecutive Fibonacci numbers \cite{berstel_fib} and the analogous freedom for the general Ostrowski representations \cite{efgms}. In the last mentioned paper, this was made to better explore the link between Ostrowski representations and Sturmian words which can also be constructed with the use of a continued fraction and its directive sequence. For the link between Ostrowski representations and Sturmian words, see also \cite{berthe};  for a modern definition of an Ostrowski numeration system, see \cite{a_sh}. For a general introduction to Sturmian words, see \cite{lothaire2}.

In this paper, we extend the range of possible representations in an Ostrowski-like numeration system even further to better reflect the structure of the related Sturmian word. In particular, this allow us to describe occurrences of palindromes in a Sturmian word by representations of their ends. Note that palindromes in Sturmian words have been extensively studied, and even their occurrences were completely described by Glen \cite{glen}. However, their relation to our numeration system catches some internal structure and in particular, allows to prove for Sturmian words the 2013 conjecture by Puzynina, Zamboni and the author \cite{fpz} which can be formulated as follows.

The {\it palindromic length} of a finite word $u$ is the minimal number $Q$ of palindromes $P_1,\ldots,P_Q$ such that $u=P_1\cdots P_Q$. 

\begin{conjecture}\label{c:main}
In every infinite word which is not ultimately periodic, the palindromic length of factors (version: of prefixes) is unbounded.
\end{conjecture}

The conjecture was stated in 2013 by Puzynina, Zamboni and the author \cite{fpz} and was proved in the same paper for the case when the infinite word is $k$-power-free for some $k$; moreover, a generalisation of the original proof works for a wider class of infinite words covering in particular fixed points of morphisms. Recently, Saarela \cite{saarela} proved the equivalence of two versions of the conjecture: the palindromic length of factors is bounded if and only if this is true for prefixes. Bucci and Richomme \cite{br} managed to prove the analogue of the conjecture for {\it greedy} palindromic lengths, but their result does not help to do it for the original statement. 

So, until now, Sturmian words remained the simplest class of words for which the conjecture was not proved. As we know from a 1991 paper by Mignosi \cite{mignosi1991}, a Sturmian word is $k$-power-free for some $k$ if and only if the directive sequence of the respective continued fraction is bounded. The case when it is unbounded does not fall into any class for which the conjecture has been proved. Moreover, computational experiments by Bucci and Richomme \cite{br} showed that the minimal length of a Sturmian factor whose palindromic length is $n$ can grow surpisingly fast. 

In this paper we use a new technique related to extended Sturmian numeration systems to prove Conjecture \ref{c:main} for every Sturmian word with an unbounded directive sequence. The word of a palindromic length greater than a given $Q$ is found explicitly as the prefix of the characteristic Sturmian word of length whose Ostrowski-like representation is of a given form. Since every Sturmian word has the same set of factors as some characteristic one, and due to the above-mentioned result by Saarela \cite{saarela}, this proves both versions of the conjecture for all Sturmian words.

We believe that first, our extension of the Ostrowski numeration system can be useful for other related problems, and second, that the developed technique can be generalized to prove Conjecture \ref{c:main} for all infinite words for which it is not yet proved, even though the proof has to be much bulkier. Moreover, new numeration systems themselves, as well as their link to palindromes, make a beautiful object for further studies.

The paper is organized as follows. In the next section, after introducing some basic notation, we define new Sturmian representations of non-negative integers. In Section \ref{s:3} we prove some properties of these representations and, inevitably, of Sturmian words. The main result of this section is Corollary \ref{c:rhobeta} stating that all valid representations of the same number can be obtained one from another by a series of elementary transformations. Section \ref{s:4} is devoted to representations of ends of palindromes and establishes relations between them (Theorem \ref{t:pr}). At last, in Section \ref{s:5}, the described properties are used to prove that the palindromic length of prefixes of characteristic Sturmian words is unbounded.

\section{Notation and Sturmian representations}
We use the notation usual in combinatorics on words; the reader is referred, for example, to \cite{lothaire2} for an introduction on it. Given a finite word $u$, we denote its length by $|u|$. The power $u^k$ means just a concatenation $u^k=\underbrace{u\cdots u}_{k}$. Symbols of finite or infinite words are denoted by $u[i]$, so that $u=u[1]u[2]\cdots$. A factor $w[i+1]w[i+2]\cdots w[j]$ of a finite or infinite word $w$, or, more precisely, its occurrence starting from the position $i+1$ of $w$,  is denoted by $w(i..j]$. In particular, for $j>0$, $w(0..j]$ is the prefix of $w$ of length $j$.

Sturmian words can be defined in many different ways discussed in detail in \cite{lothaire2}. What we use in this paper the classical construction related to a {\it directive sequence} $(d_0,d_1,\ldots,d_n,\ldots)$, where $d_i\geq 1$. Given a directive sequence, the {\it standard sequence} $(s_n)$ of words on the binary alphabet $\{a,b\}$ is defined as follows: 
\begin{equation}\label{e:def}
s_{-1}=b, s_0=a, s_{n+1}=s_n^{d_n}s_{n-1} \mbox{~for all~} n\geq 0.
\end{equation}
The word $s_n$ is called also the {\it standard word of order $n$}.

Note that to get all possible standard words, we also need to allow $d_0=0$, but due to the symmetry between $a$ and $b$, we can restrict ourselves to the case of $d_1>0$.

It is easy to see that starting from $n=0$, each $s_n$ is a prefix of $s_{n+1}$, and the lengths of $s_{n}$ strictly grow, so that there exists a right infinite word $w=\lim_{n \to \infty} s_n$. The word $w$ is called a {\it characteristic Sturmian word} associated with a sequence $(d_i)$. A word is {\it Sturmian} if its set of factors coincides with that of some characteristic Sturmian word. For a classical survey on Sturmian words and in particular characteristic Sturmian words see, e.~g., \cite{lothaire2}. 

Note that the directive sequence $(d_i)$ is closely related to the continuous fraction expansion of the {\it slope} of the Sturmian word. However, continuous fractions are not directly used in this paper. However, we will make use of the coefficients $q_n$ which are lengths of the words 
$s_n$: by the definition of $s_n$, we have
\[q_{-1}=q_0=1, q_{n+1}=d_n q_n+q_{n-1} \mbox{~for all~} n\geq 0.\]
In the {\it Ostrowski numeration system} \cite{ostr,a_sh} associated with the sequence $(d_i)$, a non-negative integer $N<q_{j+1}$ is represented as
\begin{equation}\label{e:numr}
N=\sum_{0\leq i \leq n} k_i q_i,
\end{equation}
where $0\leq k_i \leq d_i$ for $i \geq 0$, and for $i \geq 1$, if $k_i=d_i$, then $k_{i-1}=0$.
Such a representation of $N$ is unique up to leading zeros (see Theorem 3.9.1 in \cite{a_sh}). We use the notation $N=\overline{k_n \cdots k_1 k_0}[o]$. Note also that everywhere in the text, we will not distinguish representations which differ only by leading zeros.

The following lemma is a well-known direct corollary of the definition of characteristic Sturmian words and of the Ostrowski numeration system. Further in this paper we will give a proof for a more general result, named here as Corollary \ref{c:legalvalid}.
\begin{lemma}\label{l:o}
Let $w$ be the characteristic Sturmian word associated with the directive sequence $(d_i)$ and $N=\overline{k_n \cdots k_1 k_0}[o]$ in the respective Ostrowski numeration system. Then $w(0..N]=s_n^{k_n} s_{n-1}^{k_{n-1}}\cdots s_1^{k_1} s_0^{k_0}$.
\end{lemma}
\begin{example}
 {\rm 
The directive sequence $(1,1,1,\ldots)$ corresponds to the famous Fibonacci word defined by its prefixes $s_0=a$, $s_1=s_0 s_{-1}=ab$, $s_2=s_1s_0=aba$, $s_3=s_2s_1=abaab$, etc.:
\[w=abaababaabaababaababa\cdots.\]
The lengths $q_i=|s_i|$ are Fibonacci numbers, and the respective numeration system is the Fibonacci, or Zeckendorf, one: it corresponds to the greedy decomposition of a number $N$ to a sum of Fibonacci numbers $F_n$: here we start from $F_0=1$, $F_1=2$. For example, $14=13+1=F_{5}+F_0$, and thus $14=\overline{100001}[o]$. 
}
\end{example}
In several recent papers \cite{efgms,berstel_fib}, more general {\it legal} decompositions of the same type have been considered. A decomposition $N=\sum_{0\leq i \leq n} k_i q_i$ is called {\it legal} if $0\leq k_i \leq d_i$ for $i \geq 0$ (but the second restriction from the definition of the Ostrowski representation is not imposed). A number can admit several legal representations, including the Ostrowski one. A legal representation of $N$ is denoted by $N=\overline{k_n \cdots k_1 k_0}$ (without $[o]$ at the end, reserved for the Ostrowski version).

In this paper, we extend the definition of such a decomposition even more by defining general Sturmian representations of non-negative integers.

\begin{definition}
 {\rm
Given a directive sequence $(d_i)$ and the respective sequence $(s_i)$ of standard words, $|s_i|=q_i$, we call a decomposition $N=\sum_{0\leq i \leq n} k_i q_i$ {\it valid} if the prefix of length $N$ of the characteristic Sturmian word $w$ is equal to $s_n^{k_n} s_{n-1}^{k_{n-1}}\cdots s_1^{k_1} s_0^{k_0}$. A valid representation of $N$ is also denoted as $N=\overline{k_n \cdots k_1 k_0}$. We are going to discuss why it is correct to have the same notation for legal and valid representations.
}
\end{definition}

\begin{example}\label{e:1300}{\rm
As we have seen in the previous example, in the Fibonacci word, the Ostrowski representation of 14 is $14=\overline{100001}[o]$. We also have a legal representation $14=\overline{11001}$ since $14=F_4+F_3+F_0=8+5+1$. This representation is also valid, since the prefix $w(0..14]=abaababaabaaba$ of the Fibonacci word is equal to $s_4s_3s_0$: indeed,
$s_0=a$, $s_3=s_2s_1=abaab$, and $s_4=s_3s_2=abaababa$.

Now consider the representation $14=\overline{1300}$. It is valid since $w(0..14]=s_3 s_2^3 = (abaab)(aba)^3$, but not legal since its digit $b_2=3$ is greater than $d_2=1$.}
\end{example}

\section{Properties of Sturmian words and valid representations}\label{s:3}
Throughout this section, we consider a fixed directive sequence $(d_i)$ and the standard words $s_i$ (and other notions) associated with it.

In particular, we need the following classical properties \cite{dlm,lothaire2}:
 for all $n\geq 0$, we have $s_ns_{n-1}\neq s_{n-1}s_n$, but these two words coincide except for the last two symbols: $s_n s_{n-1}=c_{n}xy$ and $s_{n-1}s_n=c_{n} yx$, where if $n$ is even, $x=a$ and $y=b$ and if $n$ is odd, $x=b$ and $y=a$. The words $c_{n}$ as well as all words $c_{n,j}=s_n^{j-1}c_{n}$ for $j>0$, obtained by erasing two last symbols from $s_n^j s_{n-1}$, are called {\it central} (Sturmian) words and are palindromes.

The following proposition can be found e.~g. in \cite{berstel1999}.
\begin{proposition}\label{p:ddd}
For all $n\geq 0$, we have $s_n^{d_n} s_{n-1}^{d_{n-1}} \cdots s_0^{d_0}=c_{n+1}$.
\end{proposition}

\begin{proposition}
For all $k_0,\ldots,k_n$ such that $k_i \leq d_i$, the word $s_n^{k_n} s_{n-1}^{k_{n-1}} \cdots s_0^{k_0}$ is a prefix of $c_{n+1}$. 
\end{proposition}
\noindent {\sc Proof.} For $n=0$, the statement is obvious. To proceed by induction on $n$, it is clearly sufficient to prove that if $u$ is a prefix of $c_{n}$, then $s_n^{k_n} u$ is a prefix of $c_{n+1}$. To see it, suppose first that $k_n<d_n$. In this situation, we consider $c_{n}$ as a prefix of $s_{n}s_{n-1}$ and see that $s_n^{k_n} u$ is a prefix of $s_n^{k_n+1}s_{n-1}$ which is in its turn a prefix of $s_{n+1}$, which is a prefix of $c_{n+1}$. Now suppose that $k_n=d_n$ and consider $c_{n}$ as a prefix of $s_{n-1}s_{n}$ (obtained by erasing two last symbols). We see that $s_n^{k_n} u=s_n^{d_n} u$ is a prefix of $s_n^{d_n} s_{n-1}s_n=s_{n+1}s_n$ obtained by erasing at least two last symbols, that is, it is a prefix of $c_{n+1}$. \hfill $\Box$

\medskip
The next corollary explains why we feel free to use the same notation for legal and valid representations.

\begin{corollary}\label{c:legalvalid}
Every legal representation of a number $N$ in a Sturmian representation system is valid.
\end{corollary}
\noindent {\sc Proof.} Follows immediately from the definitions, the previous proposition and the fact that for all $n$, the word $c_{n+1}$ is by the construction a prefix of the characteristic Sturmian word. \hfill $\Box$

\medskip
In particular, Corollary \ref{c:legalvalid} implies Lemma \ref{l:o}: the Ostrowski representation is valid. 

Here is yet another property of valid representations.
\begin{proposition}\label{p:d+2}
Let $\overline{k_n\cdots k_0}$ be a valid representation of a number $N$. Then $k_0\leq d_0+1$, $k_1\leq d_1+1$, and for all $i\geq 2$, we have $k_i \leq d_i+2$. 
\end{proposition}
\noindent {\sc Proof.} The restriction $k_0\leq d_0+1$ follows from the fact that by the construction, the maximal number of consecutive $a$s in $w$ is $d_0+1$. The restrictions $k_1\leq d_1+1$ and $k_i \leq d_i+2$ for $i \geq 2$ follow immediately from Lemma 3.4 of \cite{dl2002} stating that the maximal power of $s_i$ in $w$ is $2+d_i+(q_{n-1}-2)/q_n$. \hfill $\Box$

\medskip
Now let us define two basic transformations of valid representations. Consider a representation  $r=k_n\cdots k_0$, where $k_m=d_m$ and $k_{m-1}>0$ for some $m\geq 1$, so, $r=k_n\cdots k_{m+1} d_m k_{m-1} \cdots k_0$. To distinguish the representation and the number $N$ it denotes, we write $N=\overline{r}$. The {\it unbending} transformation $\rho_m$ is defined on such representations by 
$$\rho_m(k_n\cdots k_{m+1} \cdot d_m \cdot k_{m-1} \cdots k_0)=k_n\cdots (k_{m+1}+1) \cdot 0 \cdot (k_{m-1}-1)\cdots k_0.$$ 
(Here and below we sometimes put dots between digits for readability.)
In particular, for $m=n$, we have to add to the representation the new symbol $k_{n+1}=1$: $\rho_n(d_n k_{n-1}\cdots k_0)=1 \cdot 0 \cdot (k_{n-1}-1) \cdots k_0$.

The inverse {\it bending} transformation $\beta_m=\rho_m^{-1}$ for $m \geq 1$ is defined on representations with $k_{m+1}>0$, $k_m=0$ by 
$$\beta_m(k_n\cdots k_{m+1}\cdot 0 \cdot k_{m-1} \cdots k_0)=k_n\cdots (k_{m+1}-1) \cdot d_m \cdot (k_{m-1}+1) \cdots k_0.$$

\begin{proposition}\label{p:rhobeta}
 Consider a representation $N=\overline{r}$ such that $\rho_m(r)$ (respectively, $\beta_m(r)$) is well-defined for some $m\geq 1$. If this representation is valid, then $N=\overline{\rho_m(r)}$ ($N=\overline{\beta_m(r)}$) is also a valid representation of $N$.
\end{proposition}
\noindent {\sc Proof.} Consider a valid representation $r=k_n\cdots k_0$. The fact that $\rho_m$ is well-defined means that $k_m=d_m$ and $k_{m-1}>0$. By the definition of a valid representation, we have that the prefix of $w$ of length $n$ is equal to $s_n^{k_n}\cdots s_{m+1}^{k_{m+1}}s_m^{d_m} s_{m-1}^{k_{m-1}} \cdots s_0^{k_0}$. But $s_m^{d_m}s_{m-1}=s_{m+1}$, so, the same prefix is also equal to $s_n^{k_n}\cdots s_{m+1}^{k_{m+1}+1}s_{m-1}^{k_{m-1}-1} \cdots s_0^{k_0}$. The proof for $\beta_m$ is symmetric. \hfill $\Box$

\begin{example}
 {\rm
Let us start from the representation $14=\overline{1300}$ in the Fibonacci numeration system considered in Example \ref{e:1300}. Here $d_m=1$ for all $m$. We have
\[1300 \xrightarrow{\rho_3} 10200 \xrightarrow{\beta_1} 10111 \xrightarrow{\rho_2} 11001 \xrightarrow{\rho_4} 100001.\]
Due to the previous proposition, we obtain that $14=\overline{100001}$; in fact, this is the Ostrowski representation. Note that we can also invert this series of transformations:
\[100001 \xrightarrow{\beta_4} 11001 \xrightarrow{\beta_2} 10111 \xrightarrow{\rho_1} 10200 \xrightarrow{\beta_3} 1300.\]
}
\end{example}

\medskip
We are going to prove that any valid representation can be transformed to the Ostrowski one by a series of unbending and bending transformations. To do it, we need two more propositions.
\begin{proposition}
Let us consider a sequence of coefficients $k_m,\ldots,k_n$ of length $l=n-m+1$ such that $m \geq 0$, $k_m<d_m$, and for each $i$ greater than $m$, we have $k_i\leq d_i$ with $k_i=d_i$ implying $k_{i-1}=0$. Then the word $u_{l}=s_n^{k_n}\cdots s_m^{k_m}$ is a prefix of $w$ which is followed in $w$ by the word $s_m s_{m-1}$.
\end{proposition}
\noindent {\sc Proof.} Let us proceed by induction on the length $l$ of the sequence, $l=n-m+1$. If $l=0$, the prefix $u_0$ is empty, so, it is sufficient to notice that $w$ starts with $s_{n+1}s_n$. If $l=1$, we have $u_1=s_n^{k_n}$, where $k_n<d_n$, and use the above observation for $l=0$ and the fact that $s_{n+1}=s_n^{d_n}s_{n-1}$. So, $u_1$ is followed in $w$ either by $s_n s_{n-1}$ (if $k_n=d_n-1$) or by $s_ns_n$ (if $k_n<d_{n}-1$), but since $s_n$ starts with $s_{n-1}$, this gives us what we need anyway.

Now suppose that $l \geq 2$ and that the statement is proved for $l-1$ and $l-2$ (that is, for $m+1$ and $m+2$). Let us prove it for $l$ (and $m$). By the assertion, we have $k_m<d_m$. 

If $k_{m+1}<d_{m+1}$, we use the statement for $l-1$ to see that $u_{l-1}$ is followed by $s_{m+1}s_m$. When we add $s_m^{k_m}$ to $u_{l-1}$ to get $u_{l}$, we erase from $s_{m+1}s_m=s_m^{d_m}s_{m-1}s_m$ the prefix $s_m^{k_m}$, and as above, we see that what remains starts from $s_ms_{m-1}$.

If $k_{m+1}=d_{m+1}$, we by the assertion have $k_{m+2}<d_{m+2}$ and $k_m=0$. By the induction hypothesis, $u_{l-2}$ is followed by $s_{m+2}s_{m+1}=s_{m+1}^{d_{m+1}}s_m s_{m+1}$. To get $u_{l}$, we add to $u_{l-2}$ the word $s_{m+1}^{d_{m+1}}$. In $w$, it is continued by $s_ms_{m+1}$, and since $s_{m+1}$ starts with $s_{m-1}$, the statement is proved. \hfill $\Box$

\begin{proposition}\label{p:km+1}
Let $\overline{k_n\cdots k_0}$ be a valid representation of a number $N$. Suppose that for some $m \geq 0$, we have $k_{m+1}<d_{m+1}$, and for each $i$ greater than $m+1$, we have $k_i\leq d_i$ with $k_i=d_i$ implying $k_{i-1}=0$. Then $k_{m}\leq d_{m}+1$, and the equality $k_{m}= d_{m}+1$ implies that $m\geq 2$ and $k_{m-1}=0$.
\end{proposition}
\noindent {\sc Proof.} As it was proved in the previous statement, the prefix $s_n^{k_n}\cdots s_{m+1}^{k_{m+1}}$ of $w$ is followed by $s_{m+1} s_{m}$. If $m=0$, $s_{m+1}=s_1=a^{d_0}b$, meaning that $k_{m}=k_0\leq d_0$. If $m=1$, $s_2s_1=(a^{d_0}b)^{d_1}a^{d_0+1}b$, and we see that again, $k_{m}=k_1\leq d_1$. If $m \geq 2$, then $s_{m+1} s_{m}=s_{m}^{d_{m}}s_{m-1}s_{m}$. We know that $s_{m-1}s_{m}$ differs from $s_{m}s_{m-1}$ exactly by the last two symbols, whereas $s_{m-1}$ is of length at least 2 and $s_{m-1}$ is a prefix of $s_{m}$. So, $s_{m+1} s_{m}$ starts with $s_{m}^{d_{m}+1}$ but is not equal to $s_{m}^{d_{m}+1}s_{m-1}$, the word of the same length which is a prefix of $s_{m}^{d_{m}+2}$. This means exactly that $k_{m}\leq d_{m}+1$. Moreover, if $k_{m}= d_{m}+1$, the next factor of $w$ of length $q_{m-1}$ is not equal to $s_{m-1}$, which means that $k_{m-1}=0$. \hfill $\Box$

\begin{theorem}\label{t:rhobeta}
A representation of a number $N$ is valid if and only if it can be transformed to the Ostrowski representation of $N$ by a series of unbending and bending transformations.
\end{theorem}
\noindent {\sc Proof.} The Ostrowski representation is valid due to Corollary \ref{c:legalvalid}, and since unbending and bending transformations are inversible and preserve validity due to Proposition \ref{p:rhobeta}, the ``if'' part follows.

Now let us consider a valid representation $r_0=k_n\cdots k_0$ and get from it the Ostrowski representation as follows. We look for the greatest $m$ such that either $k_m=d_m$ and $k_{m-1}>0$, or $k_m>d_m$. Due to Proposition \ref{p:km+1}, the second situation is possible only if $k_m=d_m+1$, $m\geq 2$ and $k_{m-1}=0$.

So, if $k_m=d_m$ and $k_{m-1}>0$, we take $r_1=\rho_m(r_0)$, turning the symbols $k_{m+1} \cdot d_m \cdot k_{m-1}$ into $(k_{m+1}+1) \cdot  0  \cdot (k_{m-1}-1)$. If $k_m=d_m+1$, $m\geq 2$ and $k_{m-1}=0$, we take $r_1=\rho_m(\beta_{m-1}(r_0))$, turning $k_{m+1}  \cdot (d_m+1) \cdot  0 \cdot  k_{m-2}$ first to $k_{m+1}  \cdot d_m  \cdot d_{m-1}  \cdot (k_{m-2}+1)$ and then to $(k_{m+1}+1) \cdot  0  \cdot (d_{m-1}-1)  \cdot (k_{m-2}+1)$. Note that both transformations are possible since by our conditions, $k_{m+1}<d_{m+1}$, and that due to Proposition \ref{p:rhobeta}, $r_1$ is also a valid representation of $N$.

Now we repeat the same procedure and similarly get valid representations $r_2$, $r_3$ and so on, every time transforming
the leftmost fragment with $k_m=d_m$ and $k_{m-1}>0$, or with $k_m=d_m+1$, as it is described above: so, either $r_{i+1}=\rho_m(r_i)$, or $r_{i+1}=\rho_m(\beta_{m-1}(r_i))$.  As soon as there are no such fragments, we get the Ostrowski representation. It remains only to prove that the process is always finite.

Indeed, each transformation of the type $r_{i+1}=\rho_m(\beta_{m-1}(r_i))$ decreases the greatest position $m$ of the digit $k_m$ such that $k_m>d_m$. And each transformation of the type $r_{i+1}=\rho_m(r_i)$ does not increase that position and decreases the sum of digits in the representation. Thus the chain of transformations cannot be infinite, and the theorem is proved. \hfill $\Box$

\begin{corollary}\label{c:rhobeta}
Every two valid representations of the same number can be obtained from each other by a series of unbending and bending transformations.
\end{corollary}
\noindent {\sc Proof.} The statement directly follows from Theorem \ref{t:rhobeta} and the fact that every transformation is inversible: the inverse of a bending transformation is an unbending transformation and vice versa. \hfill $\Box$

\medskip
Note that a complete study of valid representations is not within the goals of this paper. Below we just give some their properties which will be useful later.
\begin{proposition}\label{p:21}
Consider a representation $N=\overline{k_n\cdots k_0}$ where $k_m\geq 2$ and $k_{m-1}\geq 1$ for some $m\geq 1$. Then for any other representation $N=\overline{b_{n'}\cdots b_0}$ obtained from it by a series of transformations $\beta_i$ and $\rho_i$ for $i<m$, we have $b_m\geq k_m-1$.
\end{proposition}
\noindent {\sc Proof.} Let us proceed by induction on $m$. If $m=1$, the statement is obvious since $\beta_i$ and $\rho_i$ for $i<1$ are not even defined. If $m=2$, we cannot apply $\beta_1$ since $k_1>0$, and the use of $\rho_1$ can only increase $k_2$.

Suppose now that $m>2$ and the statement is proved for $m-2$. To decrease $k_m$, we should first reduce $k_{m-1}$ to zero. This is possible at least if $k_{m-1}=1$ and $k_{m-2}=0$: then we have
\begin{eqnarray*}
\cdots k_m 1 0 k_{m-3} \cdots & \xrightarrow{\beta_{m-2}}& \cdots k_m 0 d_{m-2} (k_{m-3}+1) \cdots \\
&\xrightarrow{\beta_{m-1}}& \cdots (k_m-1) \cdot d_{m-1} \cdot (d_{m-2}+1) \cdot (k_{m-3}+1) \cdots
\end{eqnarray*}
(here again dots are added for the sake of readability). To decrease the digit at position $m$ further, we should first turn $d_{m-1}$ to zero, and to do it, we should first do the same with the digit at position $m-2$. But the situation at positions $m-2$ and $m-3$ falls into the induction step: we have $d_{m-2}+1 \geq 2$ and $k_{m-3}+1\geq 1$. The use of $\rho_{m-1}$ and $\rho_{m-2}$ does not help since can only invert the previously used transformations. So, further decrease of the symbol number $m$ is not possible, which was to be proved. \hfill $\Box$

\begin{proposition}\label{p:d3}
Let $r=k_n\cdots k_0$ and $b_n\cdots b_0$ be two valid representations of the same number $N$ (we may assume that their lengths are equal since it is not prohibited to add zeros at the left). Suppose that for some $m$, where $0\leq m \leq n$, we have $4 \leq k_m \leq d_m-4$. Then $|b_m-k_m|\leq 3$.
\end{proposition}
\noindent {\sc Proof.} Due to Corollary \ref{c:rhobeta}, every representation of $N$ can be obtained from $r$ by bending and unbending transformations. Since $k_m$ is not equal to 0 nor $d_m$, we cannot apply $\rho_m$ or $\beta_m$ until $k_m$ is sufficiently changed. So, we have a restriction to use only transformations $\beta_i$ and $\rho_i$ with $i>m$ or $i<m$. These two groups of transformations do not affect each other except for the change of the digit number $m$.

Suppose that we try to decrease $k_m$. For $i>m$, the only transformation which can do it is $\rho_{m+1}$ (if $k_{m+1}= d_{m+1}$). For $i<m$, we can apply $\beta_{m-1}$ if $k_{m-1}=0$. After applying these two transformations, in any order, the fragment $\cdots k_{m+2} d_{m+1} k_m 0 k_{m-2} \cdots$ is transformed into 
$$\cdots (k_{m+2}+1) \cdot 0 \cdot (k_m-2) \cdot d_{m-1} \cdot (k_{m-2}+1)\cdots.$$
We see that there is no mean to decrease the digit number $m$ by transformations with $i>m$, since there is no mean to increase the digit number $m+1$ without increasing the digit number $m$ if the digit number $m+2$ is non-zero. As for the transformations with $i<m$, we use Proposition \ref{p:21} to show that the maximal possible decrease in this situation is by one. Note that the proposition can be applied since $k_m-2\geq 2$ and $d_{m-1}\geq 1$.

So, the total possible decrease of the digit number $m$ is at most 3. The proof for the increase is symmetric. \hfill $\Box$

\begin{example}
 {\rm
The change by 3 is indeed possible. Consider the directive sequence with $d_i=1$ for $i=0,\ldots,3,5$, and $d_4>4$. Then
$$
140000 \xrightarrow{\rho_{5}} 1030000 \xrightarrow{\beta_{3}} 1021100 \xrightarrow{\beta_{1}} 1021011 \xrightarrow{\beta_{2}} 1020121 \xrightarrow{\beta_{3}} 1011221.$$
We see that the digit number 4 was decreased from 4 to 1.
}
\end{example}

\section{Palindromes and their representations}\label{s:4}
In this section, we investigate valid representations of palindromes which occur in characteristic Sturmian words. 

The main statement of the section which will be proved below is 
\begin{theorem}\label{t:pr}
Let $w$ be a characteristic Sturmian word corresponding to the directive sequence $(d_n)$, and $w(p_1..p_2]$ be a palindrome. Then there exists a valid representation $p_1=\overline{x_n\cdots x_0}$ and a number $m \leq n$ such that $x_i\leq d_i$ for all $i<m$ and $\overline{x_n\cdots x_{m+1} y_m \cdot (d_{m-1}-x_{m-1})\cdots (d_0-x_0)}$ is a valid representation of $p_2$ for some $y_m$.
\end{theorem}
\begin{example}
{\rm
Consider the palindrome $w(12..13]=w[13]=b$ in the Fibonacci word 
$$ w=abaababaabaababaababaabaab\cdots.$$
The Ostrowski representations of $12=\overline{10101}[o]$ and $13=\overline{100000}[o]$ are quite different. However, the representations $12=\overline{1201}$ and $13=\overline{1210}$ are also valid and fit the statement of the theorem for $m=1$.

Note also that the pair of representations from Theorem \ref{t:pr} is not obliged to be unique. For example, for the palindrome $w(7..9]=aa$, we have $7=\overline{1010}$ and $9=\overline{1012}=\overline{1101}$. Both representations of $9$ constitute a pair from Theorem \ref{t:pr} with the representation of $7$: for the first one, $m=0$, and for the second one, $m=2$. 
}
\end{example}

To prove the theorem, we first have to describe some properties of palindromes in Sturmian words.

Let $w$ be an infinite word and its factor $w(p_1..p_2]$ be a palindrome. If $w(p_1-1..p_2+1]$ is also a palindrome, it is called the palindromic {\it extension} of $w(p_1..p_2]$. We can continue the process until either after $p_1$ steps we get a palindrome prefix $w(0..p_2+p_1]$ of $w$, or until $w(p_1-d-1..p_2+d+1]$ is not a palindrome for some $d<p_1$. In both cases, we speak of the {\it maximal (palindromic) extension} $w(0..p_2+p_1]$ or $w(p_1-d..p_2+d]$ of the occurrence $w(p_1..p_2]$.

To prove the next proposition, we need to know that a factor $u$ of an infinite words $w$ over the alphabet $\{a,b\}$ is called {\it left special} if $au$ and $bu$ are both factors of $w$. Symmetrically, it is called {\it right special} if $ua$, $ub$ are factors of $w$. A word which is left and right special is called {\it bispecial}. 

The set of factors of a Sturmian word is closed under reversal (see Proposition 2.1.19 of \cite{lothaire2}). It is also known (see Subsection 2.1.3 of \cite{lothaire2}) that all prefixes of characteristic Sturmian words are left special. This immediately means that bispecial factors of a characteristic Sturmian word are exactly its prefixes which are palindromes. 

\begin{proposition}\label{p:bisp}
For every occurrence of a palindrome to a characteristic Sturmian word $w$, its maximal palindromic extension is bispecial.
\end{proposition}
\noindent {\sc Proof.} If the maximal palindromic extension is a prefix of $w$, it is bispecial as discussed above. If it is of the form $w(p_1-d..p_2+d]$ with $d<p_1$, where $w(p_1..p_2]$ is the initial palindrome, it means that its left and right extension letters are different: $w[p_1-d] \neq w[p_2+d+1]$. Since the set of factors of $w$ is closed under reversal, and $w(p_1-d..p_2+d]$ is a palindrome, it means that it is bispecial. \hfill $\Box$

Bispecial factors in a Sturmian word constructed with a given directive sequence (which are also exactly prefixes of the characteristic word that are palindromes) are completely described (see \cite{dlm,lothaire2}). They are are exactly central words defined above as follows: for each $n\geq 0$, $c_n$ is the word obtained from $s_ns_{n-1}$ by erasing two last symbols, and for each $j\geq 0$, the word $c_{n,j}$ is defined as $c_{n,j}=s_n^{j}c_{n}$, so, $c_{n,j}$ is obtained by erasing two last symbols from $s_n^{j+1} s_{n-1}$. In particular, $c_n=c_{n,0}$ for all $n$. Also, $c_0$ is the empty word, $c_{0,j}=a^j$, $c_1=a^{d_0}=c_{0,d_{0}}$, $c_{1,j}=(a^{d_0}b)^ja^{d_0}$, $c_2=c_{1,d_1}$, and so on.

The following fact is a reformulation of results of \cite{dlm,berstel1999}.
\begin{proposition}\label{p:central}
 The bispecial factors of a caracteristic Sturmian word $w$ are exactly words $c_{n,j}$, where
 $0\leq j\leq d_n$. 
All these words are palindromes and each $c_{n,j}$ except for $c_{0,0}$ which is empty and $c_{1,0}=a^{d_0}=c_{0,d_{0}}$ starts with $s_n$.
\end{proposition}
 
\begin{remark}{\rm
There is no doubt that the link between Sturmian palindromes and central words has been known to specialists since many years. However, it is difficult to find references with precisely needed statements. For example, in \cite{dlDl}, it was proved that every palindrome in a Sturmian word is a median factor of a central word, but this statement concerned words and not their occurrences. For a recent study of bispecial Sturmian words, the reader is referred to \cite{fici}.
}
\end{remark}

Now we shall use the following information on occurrences of $s_n$ to $w$. First of all, since $w$ is built as the limit of the iterative construction \eqref{e:def}, for each $n$, the word $w$ can be written as a product of blocks $s_n$ and $s_{n-1}$. Following \cite{dl2002}, we call this decomposition the {\it $n$-partition} of $w$. We also use Lemma 3.3 from the same paper by Damanik and Lenz which we reformulate as follows.

\begin{proposition}[\cite{dl2002}]\label{p:33}
Consider an occurrence $s_m=w(r..r+q_m]$ of $s_m$ to $w$, where $m \geq 0$. Then $w(0..r]$ consists of full blocks of the $m$-partition of $w$, and their sequence is completely determined by the symbol $w[r]$.
\end{proposition}

\begin{proposition}\label{p:sm}
Consider an occurrence $s_m=w(r..r+q_m]$ of $s_m$ to $w$, where $m \geq 0$. Then 
\[w(0..r]=s_n^{k_n}\cdots s_m^{k_m}\]
for some $n \geq m$ and appropriate $k_i\geq 0$.
\end{proposition}
\noindent {\sc Proof.} If $r=0$, the statement is obvious with $n=m$ and $k_m=0$. For the general case, consider the partition of $w(0..r]$ to blocks equal to $s_m$ and $s_{m-1}$ which exists due to Proposition \ref{p:33}. Suppose that it ends by $k$ occurrences of $s_m$ preceded by $s_{m-1}$; we put $k_m=k$ and use the fact that by the construction, $s_{m-1}$
appears in the $m$-partition only as the end of a block $s_m^{d_m}s_{m-1}=s_{m+1}$. So, we may pass to the $m+1$-partition of $w(0..r-k_mq_m]$ and do as before: let this partition end by $k$ occurrences of $s_{m+1}$; we put $k_{m+1}=k$ and pass to the $m+2$-partition of the remaining prefix of $w$. After a finite number of steps, we will see the empty prefix, which will terminate the procedure. \hfill $\Box$

\begin{proposition}\label{p:d-l}
 Consider an occurrence $w(p_1..p_2]$ of a palindrome to $w$ and its maximal palindromic extension $w(p_1-d..p_2+d]=c_{m,j}$. Then $w(p_1-d..p_1]=s_m^{l_m}s_{m-1}^{l_{m-1}}\cdots s_0^{l_0}$ and $w(p_1-d..p_2]=s_m^{L_m}s_{m-1}^{d_{m-1}-l_{m-1}}\cdots s_0^{d_0-l_0}$ for some $L_m\geq l_m$ and for some $0\leq l_i\leq d_i$ for $i=0,\ldots,m-1$.
\end{proposition}
\noindent {\sc Proof.} First of all, $c_{m,j}$ is a prefix of $w$, and so do its prefixes $w(p_1-d..p_1]$ and $w(p_1-d..p_2]$. Let $l_m$ be the maximal power of $s_m$ such that $w(p_1-d..p_1]$ starts with it: $w(p_1-d..p_1]=s_m^{l_m}u$. Note that $l_m\leq j \leq d_m$ since $d$, the distance before the beginning of the initial palindrome, is less than $|c_{m,j}|/2$. Here $u$ is a prefix of $s_m$ and thus is can be represented as $u= s_{m-1}^{l_{m-1}}\cdots s_0^{l_0}$, where $\overline{l_{m-1}\cdots l_0}$ is the Ostrowski representation of $|u|$. So, $d=\overline{l_ml_{m-1}\cdots l_0}$, which is a legal (and valid) representation. It means that 
\[d=l_mq_m+l_{m-1}q_{m-1}+\cdots + l_0q_0.\]
Due to Proposition \ref{p:ddd}, we know that $c_m=s_{m-1}^{d_{m-1}}  \cdots s_0^{d_0}$. So, $c_{m,j}=s_m^j s_{m-1}^{d_{m-1}}  \cdots s_0^{d_0}$, and in particular,
\[|c_{m,j}|=jq_m+d_{m-1}q_{m-1}+\cdots +d_0q_0.\]
The length of $w(p_1-d..p_2]$ is $|c_{m,j}|-d$. Since $l_n \leq j$ and $l_i\leq d_i$ for $i<m$, we can just subtract the two previous equalities to get
\[|w(p_1-d..p_2]|=|c_{m,j}|-d=(j-l_m)q_m+ (d_m-l_m)q_{m-1}+\cdots +(d_0-l_0)q_0.\]
So, $|w(p_1-d..p_2]|=\overline{(j-l_m)(d_m-l_m)\cdots (d_0-l_0)}$ is a legal representation. Due to Corollary \ref{c:legalvalid}, this representation is also valid. Since $w(p_1-d..p_2]$ is a prefix of $w$, this means exactly that 
$$w(p_1-d..p_2]=s_m^{j-l_m}s_{m-1}^{d_{m-1}-l_{m-1}}\cdots s_0^{d_0-l_0}.$$
It remains to set $L_m=j-l_m$. \hfill $\Box$

\medskip
\noindent {\sc Proof of Theorem \ref{t:pr}.} Given a non-empty palindrome $w(p_1..p_2]$, consider its maximal palindromic extension $w(p_1-d..p_2+d]=c_{m,j}$. The word $c_{m,j}$ is not empty, so, the only case when it does not start with $s_m$ is $c_{1,0}=a^{d_0}$. In this case, we consider the same word $a^{d_0}$ as $c_{0,d_0}$ with $m=0$. So, we may assume that $c_{m,j}$ starts with $s_m$ and use Proposition \ref{p:sm} to see that $w(0..p_1-d]=s_n^{k_n}\cdots s_m^{k_m}$
for some $n \geq m$ and $k_i\geq 0$. We can also apply Proposition \ref{p:d-l} and see that $w(p_1-d..p_1]=s_m^{l_m}s_{m-1}^{l_{m-1}}\cdots s_0^{l_0}$ and $w(p_1-d..p_2]=s_m^{L_m}s_{m-1}^{d_{m-1}-l_{m-1}}\cdots s_0^{d_0-l_0}$. Taking the concatenation, we see that
\[w(0..p_1]=s_n^{k_n}\cdots s_m^{k_m+l_m}s_{m-1}^{l_{m-1}}\cdots s_0^{l_0}\]
and 
\[w(0..p_2]=s_n^{k_n}\cdots s_m^{k_m+L_m}s_{m-1}^{d_{m-1}-l_{m-1}}\cdots s_0^{d_0-l_0}.\]
Setting $x_i=k_i$ for $i>m$, $x_i=l_i$ for $i<m$, $x_m=k_m+l_m$  and $y_m=k_m+L_m$, and using the definition of a valid representation, we get the statement of Theorem \ref{t:pr}. \hfill $\Box$

\section{Decompositions to palindromes}\label{s:5}
In this section, we use the tools developed above to prove for Sturmian words the conjecture on the decomposition to palindromes stated in \cite{fpz}. Note that on $k$-power-free infinite words, the conjecture was proved in the same paper where is was stated. A Sturmian word is $k$-power-free for some $k$ if and only if the directive sequence $(d_i)$ is bounded \cite{mignosi1991,berstel1999}. So, it remains to prove
\begin{theorem}\label{t:main}
Consider a characteristic Sturmian word $w$ with an unbounded directive sequence $(d_i)$. Then for all $Q>0$ there exists a prefix of $w$ which cannot be decomposed to a concatenation of $Q$ palindromes.
\end{theorem}

To prove the theorem, let us first introduce another notion. Given a valid representation $r=x_n\cdots x_0$, let us denote by $z_m(r)$ the distance between $x_m$ and 0 or $d_m$:
\[z_m(r)=\min(x_m,|d_m-x_m|).\]
Note that the transformations $\rho_m$ and $\beta_m$ can be applied only to representations $r$ with $z_m(r)=0$, and do not change $z_m(r)$. Note also that Theorem \ref{t:pr} has the following immediate corollary.
\begin{corollary}\label{c:pal2repr}
Let $w(p_1..p_2]$ be a palindrome. Then there exist valid representations $r_1$ of $p_1$ and $r_2$ of $p_2$ such that $z_i(r_1)=z_i(r_2)$ for all $i$ except for one value $i=m$.
\end{corollary}

With this new notion, we state the following corollary of Proposition \ref{p:d3}.
\begin{proposition}\label{p:zd}
Let $r_1$ and $r_2$ be two valid representations of the same number $N=\overline{r_1}=\overline{r_2}$. Then for all $m$, we have $|z_m(r_1)-z_m(r_2)|\leq 3$.
\end{proposition}
\noindent {\sc Proof.} Proposition \ref{p:d3} means that the statement is true for $z_m(r_1)\geq 4$. Since the $n$th symbol of $r_2$ is bounded by $d_m+2$ by Proposition \ref{p:d+2}, Proposition \ref{p:d3} also implies that if $z_m(r_1)=0$, we cannot have $z_m(r_2)\geq 4$, completing the proof. \hfill $\Box$

\medskip
\noindent {\sc Proof of Theorem \ref{t:main}.} Since $(d_i)$ is unbounded, for any given $Q>0$, we can find $Q+1$ values $m_0,\cdots,m_Q$ such that $d_{m_i}\geq 6Q+2$ for each $i$. Consider $N=\overline{x_n\cdots x_0}$, where $x_j=3Q+1$ for $n=m_i$, $i=0,\ldots,Q$, and $x_j=0$ otherwise. This representation of $N$ is legal and thus valid due to Corollary \ref{c:legalvalid}.

Now suppose that the prefix $w(0..N]$ can be represented as a concatenation of at most $Q$ palindromes, that is, there exists a sequence
\[0=p_0\leq p_1 \leq p_2 \cdots \leq p_Q=N  \]
such that for each $k=0,\ldots,Q-1$, the word $w(p_k..p_{k+1}]$ is a palindrome.

Due to Corollary \ref{c:pal2repr}, for eack $k=0,\ldots,Q-1$, there exist representations $r_{k,1}$ of $p_k$ and $r_{k+1,2}$ of $p_{k+1}$ such that $z_i(r_{k,1})$ and $z_i(r_{k+1,2})$ differ at most for one index which we denote by $i=i_k$. Note also that 0 admits essentially only one valid representation consisting of zeros (or empty, if we get rid of leading zeros). For the sake of completeness, we denote the representation $x_n\cdots x_0$ of $N$ by $r_{Q,1}$.

Due to Proposition \ref{p:zd}, for each digit $m$, we have $|z_m(r_{k,1})-z_m(r_{k,2})|\leq 3$. So, the representation $N=\overline{r_{Q,1}}=\overline{x_n\cdots x_0}$, containing $Q+1$ digits $m_i$ equal to $3Q+1$, with the distance $z_{m_i}(r_{Q,1})=3Q+1$, is obtained from the representation $0=\overline{0\cdots 0}$ by a series of $Q$ steps. At each step, first a palindrome number $k$ (here $k=0,\ldots,Q-1$) may sufficiently change exactly one distance $z_{i_k}(r)$ when we pass from the $r_{k,1}$ to $r_{k+1,2}$. Then each distance $z_i(r)$ changes at most by 3 when we pass from $r_{k+1,2}$ to $r_{k+1,1}$ (and pass from $k$ to $k+1$).

We see that after $Q$ steps of this kind, we can only have $Q$ digits in the representation such that the distance $z$ for them is greater than $3Q$. At the same time, in the starting representation $N=\overline{x_n\cdots x_0}$ such digits are $Q+1$. A contradiction. \hfill $\Box$

\medskip
\begin{remark}
 {\rm 
The prefix of length $N$ of a characteristic Sturmian word is a factor of any Sturmian word of the same slope. So, due to the Saarela's result \cite{saarela} stating the equivalence of the prefix and the factor versions of Conjecture \ref{c:main}, we have completed the proof of the conjecture for any Sturmian word.
}
\end{remark}




\section{Acknowledgement}
I am deeply grateful to Srecko Brlek and all the organizers of WORDS 2017 whose kind invitation inspired me to return to the topic.
\section*{References}

\end{document}